\documentclass{ws-procs975x65}
\usepackage{graphicx}

\def\beq{\begin{equation}}
\def\eeq{\end{equation}}

\begin{document}

\title{Linking multipole vectors and pseudoentropies for CMB analysis}
\author{Marvin Pinkwart$^{1*}$, Peter Schupp$^1$, Dominik J. Schwarz$^2$}

\address{$^1$Department of Physics and Earth Sciences, Jacobs University, 28759 Bremen, Germany\\
$^2$Fakult\"{a}t f\"{u}r Physik, Universit\"{a}t Bielefeld, 33501 Bielefeld, Germany\\
$^*$E-mail: m.pinkwart@jacobs-university.de
}

\begin{abstract}
Multipole vectors and pseudoentropies provide powerful tools for a numerically fast and vivid investigation of possible statistically anisotropic, resp.~non-Gaussian signs in CMB temperature fluctuations. After reviewing and linking these two conceptions we compare their application to data analysis using the Planck 2015 NILC full sky map. 
\end{abstract}


\bodymatter


\section{Motivation}
Over the past 50 years the cosmic microwave background (CMB) has become the main source of information about the early universe. 
It displays a nearly isotropic black body with an order $10^{-3}$ dipole modulation due to peculiar motion and order $10^{-5}$ anisotropies which depict energy density fluctuations on the last scattering surface (LSS). Motivated by the simplest inflationary scenarios and linear perturbation theory these anisotropies are commonly assumed to be statistically isotropic and Gaussian, which means that if one decomposes the relative temperature fluctuations on the celestial sphere into spherical harmonics
\beq
\delta T (\theta,\phi) = \sum_{l=1}^{\infty}\sum_{m=-l}^l a_{lm}Y_{lm}(\theta,\phi),
\eeq
the joint distribution of the spherical harmonic coefficients $a_{lm}$ for a given multipole number $l$ should be isotropic
\beq
p(\hat{R}\{ a_{lm} \}) = p(\{ a_{lm} \}),
\eeq
where the rotation acts on the $a_{lm}$ via Wigner symbols, and Gaussian
\beq
p(\{ a_{lm} \}) = \mathcal{N} \exp\left( -\frac{1}{2} \sum_{mm'}a_{lm'}^* \mathcal{D}_{l,mm'} a_{lm} \right).
\eeq
Since the one-point function of the $a_{lm}$ vanishes, Gaussianity implies that all information about $\delta T$ is encoded in the two-point functions $\langle a_{lm}a_{lm'} \rangle = \mathcal{D}_{l,mm'}^{-1}$ and statistical isotropy further reduces the degrees of freedom to one per multipole number by diagonalizing the covariance $\mathcal{D}_{l,mm'}^{-1} = C_l \delta_{mm'}$. This yields the power spectrum which is commonly used as the main source of information about the CMB. Hence, the two basic assumptions reduce the real degrees of freedom per multipole from $2l+1$ to one. If one or both of the assumptions are relaxed, one needs to find additional data compressing measures which are easier and can be interpreted more directly and vividly than the $a_{lm}$. In this contribution we describe and link two approaches to find such measures which are furthermore complementary to the power spectrum in information content. The first approach concerns multipole vectors whose use is motivated by large scale anisotropy anomalies that have been found recently and the second approach introduces pseudoentropy measures on the space of spin states associated to temperature fluctuation fields which provide a non brute-force method to investigate non-Gaussianities and are motivated by the analogy to statistical mechanics where the large number of microscopic degrees of freedom is essentially captured by a few macroscopic quantities such as entropy .

\section{Multipole vectors and pseudoentropies}

Multipole vectors (MPVs) have been introduced to CMB data analysis in Ref.~\refcite{MultiCopi}. and were used to identify one of the three basic ``atoms" of large scale isotropy anomalies. It was found that the quadrupole and octupole are correlated, that the octupole is unusually planar and that both quadrupole and octupole are unusually orthogonal to the Ecliptic Plane and aligned with the Solar Dipole, see the review in Ref.~\refcite{Schwarz2015}. Contrary to the behavior of spherical harmonic coefficients, MPVs rotate rigidly with the celestial sphere and therefore put themselves forward as the correct basic constituents of measures of statistical anisotropy. Let $\vec{e} = (e^1,e^2,e^3)^T = \vec{e}(\theta,\phi)$ be a unit vector, then the basic idea is to observe that the $l$-multipole-part of $\delta T$ is a homogeneous polynomial of degree $l$ and to decompose it according to
\beq
\left( \delta T \right)_l(\vec{e}(\theta,\phi)) = A^{(l)}\left[ v_{i_1}^{(l,1)} \cdots v_{i_l}^{(l,l)} \right] \left[ e^{i_1} \cdots e^{i_l} \right] = B^{(l)}\prod_{i=1}^l \vec{e} \cdot \vec{v}^{(l,i)} + F^{(l)}(\theta,\phi),
\eeq
where $A^{(l)}$ and $B^{(l)}$ are real constants, $F^{(l)}$ is a polynomial of degree $l-2$ and Einstein's summation convention was used.\cite{KatzWeeks} The $\vec{v}^{(i,l)}$ are the MPVs and contain all information content of the temperature map up to a constant amplitude. Since the power spectrum is contained within this amplitude, the information content of the MPVs lies in the complement of the space of $C_l$-information. Hence, MPVs provide an interesting tool to test if this subspace is non-empty, i.e. if the temperature map is statistically anisotropic. In fact, as shown in Ref.~\refcite{Dennis2007} the statistical behavior of MPVs is the same for the larger class of all completely random sets of $a_{lm}$, i.e. $p(\{ a_{lm} \}) = f(\sum_m |a_{lm}|^2)$, which contains the standard $\Lambda$CDM case but also some isotropic non-Gaussian maps. Note, that the $l$-multipole part of $\delta T$ can also be viewed as a normalized spin-$l$ state 
\beq
|\delta T \rangle_l = \sum_{m} \widetilde{a}_{lm}|lm\rangle
\eeq
up to an amplitude which contains the information about $C_l$, where the $a_{lm}$ have been normalized to $\widetilde{a}_{lm}$. Then the contraction of Bloch coherent states with such a temperature state and stereographic projection yield a complex polynomial
\beq
\label{eq:polynomial}
_l \langle \Omega|\delta T\rangle_l \sim \sum_{m}\sqrt{\binom{2l}{l+m}}\widetilde{a}_{lm}z^{l+m} \sim \prod_{i=1}^{l}(z-z^{(l,i)})\left(z+\frac{1}{(z^{(l,i)})^*}\right),
\eeq
whose $2l$ zeros are the stereographic projections $z^{(l,i)}$ of the MPVs $\vec{v}^{(l,i)}$ and their antipodes.\cite{Schupp} Bloch coherent states are given by the rotation of the highest spin state $|\Omega\rangle_l = \hat{R}(\Omega)|l,l\rangle$ by angles $\Omega = (\theta,\phi)$. For a detailed review on MPV constructions see Ref.~\refcite{PinkwartI}. Now one can define the most classical quantum entropy which is named Wehrl entropy
\beq
\label{eq:wehrl}
S^{\mathrm{W}}_l=S^{\mathrm{W}}(|\delta T\rangle_l) = -\frac{(2l+1)}{4\pi}\int \! \mathrm{d}\Omega \, |_l \langle \Omega|\delta T\rangle_l|^2 \log\left( |_l \langle \Omega|\delta T\rangle_l|^2 \right),
\eeq
and which is used for CMB data analysis as a measure of randomness of temperature fluctuations.\cite{SchuppHelling} The disadvantage of the Wehrl entropy is its numerically expensive form and hence one seeks measures that approximate the Wehrl entropy reasonably well but are much easier to compute. Since the temperature map describes a pure state, the usual von Neumann entropy is trivial. The basic idea to overcome this problem is to apply a quantum channel $\Phi$, which respects isotropy, to the pure temperature state $\rho_l = |\delta T\rangle_l\langle \delta T|$ to obtain a mixed density $\rho_l^{\Phi}$ and afterwards apply the von Neumann entropy to get a mixed pseudoentropy 
\beq
\label{eq:entropy}
S^{\Phi}_l(\delta T) = -\mathrm{Tr}\left(\rho_l^{\Phi} \log(\rho_l^{\Phi})\right). 
\eeq
The term ``pseudo" shall indicate that the resulting quantity loses some properties entropies usually possess. From now on we drop this adjunct. In Ref.~\refcite{PinkwartII} we introduced two convenient choices of rotationally symmetric quantum channels:
\begin{itemlist}
\item The $j$-projection entropy uses the mixed density
\beq
\rho^{\mathrm{proj},(j)}_l = \Phi^{\mathrm{proj},(j)}(\rho_l) = \frac{2l+1}{2(l+j)+1}  \hat{P}_{l+j} \left(\rho_l \otimes 1_j \right) \hat{P}_{l+j},
\eeq
where $\hat{P}_{l+j}$ denotes the projection to the $[l+j]$-subspace of $[l] \otimes [j]$. It converges to $S^{\mathrm{W}}_l+\log\left(\frac{2(l+j)+1}{2l+1}\right)$ for $j \rightarrow \infty$, but not uniformly. The additional term does not depend on the data, but just on the numbers $l$ and $j$ and henceforth in data analysis this term does not need to be taken into account. Therefore one may say that in the realm of data analysis the projection entropy converges to the Wehrl entropy.
\item The angular entropy uses the mixed density
\beq
\label{eq:angular}
\rho^{\mathrm{ang}}_l = \Phi^{\mathrm{ang}}(\rho_l) = \frac{1}{l(l+1)}\sum_{i=1}^3 \hat{L}_i \rho_l \hat{L}_i^\dagger,
\eeq
where the $L_i$ are the angular momentum operators acting on spin space. For $l=2$ it measures the repulsion of MPVs and shows the exact same behavior as the Wehrl entropy once an overall shift and dilation in absolute value has been accounted for. Let $\epsilon$ be the squared chordal distance between the two MPVs on a sphere of radius $1/2$, then the angular and Wehrl entropy can be expressed as
\begin{align}
\label{eq:angMPV}
S^{\mathrm{ang}}_2 (\epsilon) &= -\frac{c(\epsilon)}{2}\left( (1-\epsilon)^2\log\left(1-\epsilon\right)^2 + \epsilon^2 \log\left( \epsilon^2 \right) \right) - \log\left( \frac{c(\epsilon)}{2}\right) \\
\label{eq:wehrlMPV}
S^{\mathrm{W}}_2 (\epsilon) &= c(\epsilon) - \log(c(\epsilon)) + \frac{32}{15} - \log(6),
\end{align}
where $c(\epsilon):=(1-(1-\epsilon))^{-1}$. Motivated by this result, the convergence of the projection to the Wehrl entropy and the numerically similar behavior, we propose that the similarity of Wehrl and angular entropy as well as the interpretation as a measure of repulsion between MPVs should be valid on all scales, at least approximately. Since repulsive behavior of zeros of random polynomials is connected to the degree of Gaussianity of the polynomial, we consider the angular entropy as a simple measure for deviations from Gaussianity, at least as long the deviation causes an exit from the space of completely random distributions. Furthermore, we showed that the probability distribution of the angular entropy is highly sensitive to ``small" deviations from Gaussianity, in the sense that if one replaces $i$ MPVs from a Gaussian map by uniformly distributed unit vectors --- which constitutes a deviation from Gaussianity without violating isotropy --- the shift in distribution is sizable and the largest portion of shift is obtained from the first replaced vector.
\end{itemlist}
Since all entropy measures use the normalized temperature state, their information content is complementary to the $C_l$, which is also the case for MPVs. Using the latter to build vivid scalar quantities yields additional measures, which we interpret as measures of statistical anisotropy. In Ref.~\refcite{PinkwartI} we used the following statistics:
\begin{itemlist}
\item The internal/inner statistics
\begin{align}
\label{eq:int_align}
S^{||}(l) &= \frac{2}{l(l-1)} \sum_{i<j}|\vec{v}^{(l,i)}\cdot\vec{v}^{(l,j)}|\\
S^{v}(l) &= \frac{6}{l(l-1)(l-2)}\sum_{i<j<k}\left|\left(\vec{v}^{(l,i)}\times\vec{v}^{(l,j)}\right)\cdot \vec{v}^{(l,k)}\right|
\end{align}
measure internal alignment and non-planarity of multipoles. In structure and interpretation these statistics share some similarities with the entropy measures.
\item The external/outer statistics
\begin{align}
\label{eq:ext_align}
S^{||}_{\vec{D}}(l) &= \frac{2}{l(l-1)} \sum_{i}|\vec{v}^{(l,i)}\cdot\vec{D}|\\
S^{v}_{\vec{D}}(l) &= \frac{6}{l(l-1)(l-2)}\sum_{i<j<k}\left|\left(\vec{v}^{(l,i)}\times\vec{v}^{(l,j)}\right)\cdot \vec{D}\right|,
\end{align}
where $\vec{D}$ denotes some given outer physically motivated direction, measure outer directional or planar influences on the data.
\end{itemlist}

\section{Application to data analysis}

\begin{figure}
\begin{center}
\includegraphics[width=\textwidth]{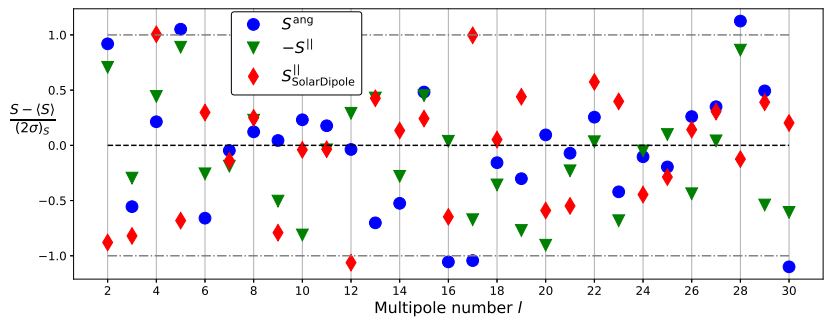}
\caption{Comparison of MPV statistics and angular entropy via the deviation of NILC 2015 full sky data from isotropic, Gaussian expectation in units of $2\sigma$-deviation calculated with $10^4$ (entropy) and $10^3$ (MPVs) ensembles on the range $l\in[2,30]$.}
\label{fig:comp}
\end{center}
\end{figure}

Both methods have been applied to Planck 2015 and Planck 2018 full sky maps in order to compare them to statistically isotropic and Gaussian random maps and investigate if any signs of possible violations appear on multipole numbers up to $50$ (MPVs), resp.~$1000$ (entropies).\cite{PinkwartI,PinkwartII} In Fig.~\ref{fig:comp} we compare $S^{\mathrm{ang}}$ \eqref{eq:entropy}\eqref{eq:angular}, -$S^{||}$\eqref{eq:int_align}, and $S^{||}_{\vec{D}}$ \eqref{eq:ext_align} with the Solar Dipole as a given physical direction $\vec{D}$ using the NILC 2015 full sky map.
We use the negative of $S^{||}$ because it allows for an interpretation as a measure of non-alignment which is related to the interpretation of $S^{\mathrm{ang}}$ as a measure of repulsion. One observes that the general behavior of both $S^{\mathrm{ang}}$ and $S^{||}$ as a function of $l$ approximately agrees while unlikely multipoles are much more pronounced in $S^{\mathrm{ang}}$. $S^{||}_{\vec{D}}$ clearly deviates from the others and the clustering of low-likelihood multipoles at $l\in[2,5]$ reproduces the corresponding atom of CMB anomalies. Note that for the sake of clarity $1/3\sigma$-boundaries are not shown due to their different relative distances among the statistics, but note that $1\sigma \neq \frac{1}{2}(2\sigma)$.
\\
Eventually we summarize further results that were obtained in Refs.~\refcite{PinkwartI,PinkwartII}. With the MPV statistics the behavior of three full sky maps shows up to be similar even without applying masks, and no global anisotropies could be identified on $l \in [2,50]$. After masking, the behavior of SEVEM approaches that of the other maps. The direction that is visible in the data most is the Solar Dipole, which appears in the large scale anomalies and via a clustering of $1\sigma$ (anti-)alignments around $l=20$. Using the entropies the findings on masking, non-masked SEVEM, similarity between the other maps and overall normal behavior agree with those obtained with MPV statistics, but the entropy shows above $2/3\sigma$ unusual values at $l=5/28$, which are curious because their values are higher than expected. It is unclear which kind of physical temperature map can result in too high entropies. The large-scale anomalies are visible when applying the range angular entropy, which takes into account ranges and possible correlations of multipoles. A comparison between 2015 and 2018 data revealed that the change in NILC and SMICA is negligible while unmasked SEVEM has been enhanced. The angular entropy can also be used on very small angular scales with moderate computational expense.

\section*{Conclusions}
MPV statistics yield vivid measures for identifying possible statistical anisotropies, possibly induced by physical influences measured via effects of outer directions. On the other hand our entropy methods yield fast measures for non-Gaussianity which allow for investigation down to small angular scales $l \geq1000$. Both methods are mathematically related, see Eqs.~\eqref{eq:polynomial}, \eqref{eq:angMPV} and \eqref{eq:wehrlMPV}, and yield information content that is complementary to that of $C_l$. Note that in the future effects of inhomogeneous noise and masking should be taken into account more thoroughly. Our methods need full sky data but using e.g. spherical cap harmonics or other local methods one could make more sense of the statistics when applying a mask. An interesting future task would also be to redo the analysis using full focal plane simulations. Finally an efficient measure for unlikeliness when picking out ranges of random size needs to be worked out.

\section*{Acknowledgments}
We acknowledge financial support by the DFG RTG 1620 Models of Gravity.
\bibliographystyle{ws-procs975x65}
\bibliography{ws-pro-sample}

\end{document}